\documentstyle{elsart}
\newcommand{\dd}{{\rm d\ }}
\begin{document}
\begin{frontmatter}

\title{%        %You can use \\ for explicit line-break
Space-time Evolution of Quark-Gluon 
Plasma Fluid with \\First Order Phase Transition
}

\author[TWC]{ Shin { Muroya}\thanksref{muroya} } and
\author[Waseda]{Chiho { Nonaka}\thanksref{nonaka}}

\address[TWC]{Tokuyama Women's College, Tokuyama, Yamaguchi, 745 Japan}
\address[Waseda]{Department of Physics, Waseda University, Shinjuku-ku, 
Tokyo, 169 Japan}

\thanks[muroya]{E-mail : muroya@yukawa.kyoto-u.ac.jp} 
\thanks[nonaka]{E-mail : 696l0906@mn.waseda.ac.jp}
\begin{abstract}
%{%         %this abstract is neglected when [addenda] or [errata]
%Write your ABSTRACT here.
We numerically solve the (3+1)-dimensional relativistic 
hydrodynamical equation 
with a bag-model equation of state, which is one of the simplest models 
for the first order phase transition.  
Based on the numerical solution, we discuss the space-time evolution of the 
hot fluid produced in ultra relativistic nuclear collisions.  Especially, 
the space-time structure of the two-phase co-existing region 
during the quark-gluon plasma-hadron phase 
transition is analyzed in detail. Finally, we investigate the change of 
the mass spectra of $J/\psi$, ${\eta}_{\rm c}$ and ${\phi}$ as possible 
experimental signatures of the phase transition.


\end{abstract}

\end{frontmatter}

\section{Introduction}
%Start your paper from here.

The QCD phase transition and quark-gluon plasma is currently one of 
the hottest topics in  high energy nuclear physics.\cite{QM}
The recent high performance Lattice simulation suggests that the order of 
the QCD phase transition in the real world, i.e. 2 massless quarks and 
1 massive quark case, is first order.\cite{Kanaya}
If the QCD phase transition is of the first order,
the latent heat will be  released at the phase transition 
temperature $T_{\rm c}$ and an isothermal  
Quark-Gluon Plasma(QGP)-hadron co-existing region will appear 
in the case that the change of the whole system 
is slower than the microscopic relaxation time. 
If such a scenario will be realized, the space time 
volume of the $T=T_{\rm c}$ region can become significantly 
large, 
and physical quantities which are characteristic to the phase 
transition temperature $T_{\rm c}$ will 
be available as the experimental signatures of the phase transition.
The purpose  of this paper is to investigate how large a region with 
$T=T_{\rm c}$ will be produced at the CERN SPS experiment and 
how clearly the hadronic mass shift at finite temperature
can work as possible signatures of the phase transition.

As an equation of the state, we here adopt the bag model 
which is one of the simplest models for the first order phase 
transition.  The hydrodynamical model for the QGP fluid has 
already been discussed in many papers since Bjorken first 
introduced the simple scaling model in 1+1 dimensional 
expansion\cite{Hydro}. We discussed 1+1 dimensional expansion of 
the viscous fluid \cite{Date1} and 3+1 dimensional expansion with 
phase transition\cite{Akase}. Based on the hydrodynamical model, we 
also discussed electromagnetic probe of QGP\cite{Hirano}. 
As concerns the 
treatment of the first order phase 
transition, Alam et al.\ and Kajantie et al.\ \cite{Alam}
focused their discussion only on the  
transverse expansion based on the scaling ansatz. 
Sollfrank et al.\ \cite{Sollfrank}
discussed a smeared equation of state with a small temperature 
width. In order to focus our discussion to the $T=T_{\rm c}$ region, 
we solved 3+1 dimensional expansion keeping an equation of state 
with the exact first order phase transition.

Experimentally, at SPS energy, net baryon number is not 
negligible even in the central region.  In order to take correctly account
of the baryon number conservation law, we must solve the coupled equations 
of hydrodynamics and baryon current conservation\cite{Ishii}.
However, if we adopt baryon number independent of an equation of state, 
 the dynamics of the four velocity is free from the 
baryon conservation law and 
the role of the chemical potential is  only for the calculation of 
the baryon number at the final stage\cite{Arbex}.  
Therefore, we may say that the baryon free fluid 
model works well enough in the discussion of space time evolution of 
the fluid with the simple bag model.

The mass shift of the hadrons in the hot medium is 
believed to be a promising detectable signature of the high temperature 
region.  Hioki et al.\ 
proposed $J/\psi$ and $\eta_{c}$ \cite{Hioki} and Asakawa et al.\  
proposed $\phi$ as the possible probe\cite{Asakawa}.
Following their discussions, we apply our hydrodynamical model to the 
evaluation of the mass spectrum of these particles.

\section{Hydrodynamical Equation}

Assuming cylindrical symmetry to the system, we introduce 
cylindrical coordinate $\tau$, $\eta$, $r$  and $\phi$ 
instead of the usual Cartesian
coordinate $x$, $y$, $z$ and $t$, where,
\begin{eqnarray}
t&=&\tau\cosh \eta, \nonumber \\
z&=&\tau\sinh \eta, \nonumber \\
x&=&r \cos \phi, \nonumber \\
y&=&r \sin \phi.
\end{eqnarray}
The four velocity $U^{\mu}$ is given as,
\begin{eqnarray}
U^{t}&=&U^{\tau}\cosh \eta +U^{\eta}\sinh \eta, \nonumber \\
U^{z}&=&U^{\tau}\sinh \eta +U^{\eta}\cosh \eta, \nonumber \\
U^{x}&=&U^{r}\cos \phi+U^{\phi}\sin \phi, \nonumber \\
U^{y}&=&U^{r}\sin \phi-U^{\phi}\cos \phi.
\end{eqnarray}
Taking into account $U^{\mu}U_{\mu}=1$ and cylindrical symmetry,
$U^{\mu}$ can be represented with two variables $Y_{\rm L}$ and 
$Y_{\rm T}$,
\begin{eqnarray}
U^{\tau}&=&\cosh Y_{\rm T} \cosh(Y_{\rm L}-\eta), \nonumber \\
U^{\eta}&=&\cosh Y_{\rm T} \sinh(Y_{\rm L}-\eta), \nonumber \\
U^{r}&=&\sinh Y_{\rm T}, \nonumber \\
U^{\phi}&=&0.
\end{eqnarray}
As is well known, the relativistic hydrodynamical equation is 
given as,
\begin{equation}
\partial_{\mu}T^{\mu\nu} = 0,
\end{equation}
and the energy momentum tensor of perfect fluid is 
given by,
\begin{equation}
T^{\mu\nu} = EU^{\mu}U^{\nu}-P(g^{\mu\nu}-U^{\mu}U^{\nu}).
\end{equation}
where $E$ and $P$ are energy density and pressure, respectively.
Usually, thermodynamical quantities such as $E$ and $P$ 
are treated as local quantities through 
temperature $T=T(x^{\mu})$.

In order to solve the above hydrodynamical equation, 
we need an equation of state.
In this paper, we adopt the bag model equation of state as 
the simplest model for 
the QCD phase transition of first order. In the bag model, thermodynamical
quantities below the phase transition temperature correspond to  those of
a massive hadronic gas. On the other hand, in the high temperature phase, 
thermodynamical quantities are given by those of massless QGP gas with
bag constant $B$.  In this paper, we assumed pions and kaons in the hadronic 
phase and u-, d-, s-quarks and gluons in the QGP phase.  Putting the phase 
transition temperature, $T_{\rm c}=160$ MeV, into the condition of 
pressure continuity, 
we can obtain the bag constant as $B= 412$ MeV/fm$^3$. 

At the phase transition temperature $T_{\rm c}$, pressure
$P$ is continuous but other quantities such as energy density $E$ and 
entropy density $S$ have discontinuity as a function of temperature.
We parameterize  these quantities during the phase transition by using 
the volume fraction $\lambda$ \cite{Hioki}.
We assume that thermodynamical quantities at $T = T_{\rm c}$ are 
the functions of the space-time point through $\lambda(x^{\mu})$ and given as,
\begin{eqnarray}
E(\lambda)&=&\lambda E_{\rm qgp}(T_{\rm c}) + 
(1 - \lambda) E_{\rm had}(T_{\rm c}), \nonumber \\
S(\lambda)&=&\lambda S_{\rm qgp}(T_{\rm c}) + 
(1 - \lambda) S_{\rm had}(T_{\rm c}),
\end{eqnarray}
where $ 0 \le \lambda(x^{\mu}) \le 1 $. 
This parameterization enables us not only to solve the hydrodynamical equation 
easily but also to estimate hadronic fraction in the phase 
transition region for later use.

\section{Space-time evolution of co-existing region}

Assuming that local equilibrium is achieved at 1 fm later than the collision 
instance, we give initial conditions of the hydrodynamical model on 
the $\tau = \tau _{0}=$ 1 fm hypersurface.  
Initial entropy density distribution is given as,

\begin{equation}
S(\tau_0,\eta)=S(T_0)\exp\bigl(-{\frac{(\vert \eta \vert-\eta_0)^2}
{2 \cdot{\sigma_\eta}^2}} \theta(\vert \eta \vert-\eta_0)
-{\frac{(r-r_0)^2}{2 \cdot{\sigma_r}^2}} \theta(r-r_0)\bigr)
\label{(3)},
\end{equation}
and parameters we used in this paper are summarized in table 1.

%---------------------------\\
\begin{center}
\begin{tabular}{cccccc}
\multicolumn{6}{c}{ Table 1.  Initial Parameters } \\
\hline
       & $T_{0}$ (MeV) & $\eta_{0}$ & $\sigma_{\eta}$ & $r_{0}$ (fm) & 
$\sigma_{\rm r}$ (fm) \\ \hline 
S + Au & 209 & 0.7 & 0.7 & $(32)^{(1/3)}-1.0 $ & 1.0 \\ \hline
Pb + Pb& 190 & 0.7 & 1.0& $(197)^{(1/3)}-1.0 $ & 1.0 \\
\hline

\end{tabular}
\end{center}
%---------------------------\\

Putting freeze-out temperature as $T_{\rm f} =140$ MeV for 
S+Au and $T_{\rm f} =138$ MeV for Pb+Pb, respectively,
we can reproduce hadronic distributions of
the recent experimental results. Figure 1 and fig.~2 show 
 the pseudo-rapidity distribution of charged hadrons and the transverse
momentum distribution of neutral pions, respectively. 
Data for  S+Au 200 AGeV collisions are
obtained by WA80\cite{92WA80,94WA80} and data of Pb+Pb 158 
AGeV collisions are obtained  by NA49\cite{96NA49} 
(rapidity distribution of charged hadrons) 
and WA98 \cite{WA98}
(transverse momentum distribution of neutral pions).

Figure 3 displays the space time evolution of the temperature distribution
 in the Pb+Pb case.  The isothermal region at phase transition temperature 
$T=T_{\rm c}$ appears very clearly.  The structure of temporal evolution 
does not differ essentially from the case of smooth phase transition 
in a small temperature region \cite{Akase,Sollfrank}.  By virtue of the
use of the volume fraction $\lambda$ at phase transition temperature, 
we can pick up the volume element with the temperature at just the phase 
transition point.
The temperature profile function (fig.~4) shows the significant contribution 
of the  space-time volume at $T=T_{\rm c}$.

\section{Change of the mass spectrum}

As a signal of the hot hadronic region, shifted $J/\psi$ mass-spectrum and 
$\eta_{\rm c}$ were proposed by Hioki et al.\cite{Hioki} and $\phi$ 
by Asakawa et al.\cite{Asakawa} In these papers,
authors discussed possible signals based on a scaling 
hydrodynamical model.
We apply our (3+1)-dimensional hydrodynamical model to these calculation 
and investigate numerically
 how clear the shifted hadronic mass spectra appear in the 
present CERN SPS case.

According to ref.\ \cite{Hioki}, the invariant mass distribution of leptonic pair decay is given by
$$
\frac{1}{\sigma} \frac{\dd \sigma}{\d M}
=\int^{T_{0}}_{T_{f}} \dd T \Phi(T) n(T) 
\Gamma_{l \bar{l}}(T) \delta(M-M(T)),
$$
where $\Phi$ is the temperature profile function which we have already 
calculated in the previous section.
In ref.\ \cite{Hioki}, Hioki et al. evaluated shifted mass $M(T)$ and 
decay width $ \Gamma_{l \bar{l}}(T) $ based on a c-$\bar{\rm c}$ 
potential model and we also use their result here.  Charmonium density 
$n(x^{\mu})$ was estimated by the kinetic equation in ref.\ \cite{Hioki},
and we adopt Maxwell-Boltzmann  distribution for $ n(x^{\mu}) $, that is,
$$ n(T) = 3\left(\displaystyle{\frac{MT}{2 \pi}}\right)^{\frac{3}{2}} 
{\rm exp}(-\frac{M}{T}),$$
for simplicity.  However, charmonium mass is much larger than the typical 
temperature of the fluid,
above $n(T)$ is almost constant and
final results were dominated by the temperature profile function $\Phi(T)$.
Figure 5 and 6 show the possible $J/\psi$ and $\eta_{\rm c}$ spectrum 
for the CERN SPS experiment in comparison with the results 
of ref.~\cite{Hioki}.  In fig.~5, ``critical" 
and ``thermal" correspond to the yields of 
 $T=T_{\rm c}$ and  $T_{f}<T<T_{\rm c}$, respectively.  
 ``Cold" in fig.~5 is given by,
$$
\displaystyle{\frac{1}{\sigma}} \frac{\dd \sigma}{\dd M}
= \frac{N \Gamma_{l \bar{l}}(T=0)}{\Gamma_{\rm total}} \delta(M-M(T))
$$
which corresponds to the naive thermal contribution from 
the freeze-out hypersurface.
Note that, in order to make comparison clear, the results of 
ref.~\cite{Hioki}, which are the (1+1)-dimensional calculation,
 were so normalized in fig.~5 and fig.~6 as to give 
the same ``cold" contribution to our results.
In fig.~5 and fig.~6, quantitative difference between S+Au and Pb+Pb 
is consistent with the difference in the temperature profile functions
(fig.~4).

The yield of $J/\psi$ from $T_{\rm c}$ is several orders smaller 
compared to the cold and thermal $J/\psi$.
However, the situation of $\eta_{c}$ is much better 
than $J/\psi$: in our results, 
$\eta_{c}$ from the $T=T_{\rm c}$ region is almost the same order as the 
others.  
Hence, $\eta_{c}$ will be a better signature of 
$T=T_{\rm c}$ than $J/\psi$.
These results are consistent with  ref.~\cite{Hioki}.
Because of the very small decay width of $J/\psi$ and $\eta_{c}$,
changes of the mass seem very clear and easy to detect.
On the other hand, their life time is much longer than the hot 
fire ball and the number of  particles which decay 
in the hot medium is very small.  Therefore, shifted $\eta_{c}$ and $J/\psi$ 
are hardly detectable in the present experimental situation, but in future 
experiments with high precision, these can work as good signals.

Asakawa et al.\cite{Asakawa} started their discussion from the number 
of  particles which decay in 
unit time and unit phase space which is given as,
\begin{eqnarray}
\dd N = \displaystyle{\frac{g_{\phi}}{(2\pi)^3}}
\displaystyle{\frac{1}{\gamma}}\Gamma_{\bar{l}l}(T) 
\e^{-p^{\mu}U_{\mu}/T} 
\dd^{4}x \dd^{3}p,
\label{eqn:p1}
\end{eqnarray}
where $g_{\phi}$ is the degeneracy of $\phi$ and $\gamma$ is the 
Lorentz factor 
of a moving particle.
They imported the results of the QCD sum rule for 
the decaying constant in finite temperature, $\Gamma(T)$.
From (\ref{eqn:p1}) we obtain 
\begin{eqnarray}
\frac{\dd N_{\bar{l}l}}{\dd M \dd y} =
\frac{g_{\phi}}{(2\pi)^3}
\int\!\!\!\int\frac{1}{\gamma}
\Gamma_{\bar{l}l}(T)F_{\phi}(M,m_{\phi}(T))
{\mbox e}^{-p^{\mu}U_{\mu}/T}\dd ^4x \dd p_{T}\dd p_{T}\dd \varphi,
\label{eqn:p2}
\end{eqnarray}
where $F_{\phi}(M,m_{\phi}(T))$ is the normalized smearing function to take 
account of experimental resolution. 
$F_{\phi}(M,m_{\phi}(T))$ is assumed to have a Gaussian form,
\begin{eqnarray}
F_{\phi}(M,m_{\phi}(T)) = \frac{1}{\sigma \sqrt{2\pi}}{\mbox e}^{-(M-m_
{\phi}(T))^2/2\sigma ^2},
\end{eqnarray}
where $\sigma$ is put as 10 MeV.
With use of the above formula, we evaluate the $\phi$ spectrum based on our 
hydrodynamical model.
Figure 7 shows the $\phi$ spectrum from S+Au and Pb+Pb collisions at CERN SPS
energy. The dashed line in Fig.\ 7 stands for the contribution of
$T=T_{\rm c}$ and the dotted line  for the $\phi$ from the hot hadronic region.
Quantitative difference between yields in S+Au(a) and Pb+Pb(b) are 
almost proportional to the difference in the temperature
 profile functions(fig.~4).
Comparing to S+Au,
the ratio of the peak around 959 MeV to the one around 1020 MeV is 
large in the Pb+Pb case.
 
These curves show clear difference as we expected, but the accumulated
result (solid line) does not show distinguishable double peaks. 
In the accumulated results, we can  find clear double peaks 
originating from the cold component and hot hadronic
contribution, however it would not mean directly the 
existence of the QCD phase transition.

\section{Concluding remarks}

We solved a (3+1)-dimensional hydrodynamical model with first order
phase transition model(bag model). 
We can parameterize  our hydrodynamical model so as to reproduce well 
the present
experimental hadron spectrum. Based on the numerical solution we
discussed the space-time structure of the isothermal region at phase transition
$T=T_{\rm c}$, which appears as a sharp peak in the temperature profile 
function. Applying our numerical solution, we evaluated mass spectra of
$J/\psi$, $\eta_{\rm c}$ and $\phi$ which have been proposed as 
possible signals for the existence of an isothermal $T=T_{\rm c}$ 
region.   According to our numerical investigation, 
if the experimental resolution is as good as $\sigma = 10$ MeV, we may 
have a chance to find the double-peak of the $\phi$ spectrum, but  
in the present CERN SPS Pb+Pb experiment, $T=T_{\rm c}$ peak is  
 indistinguishable from the hot 
hadronic contribution.

We are indebted to Prof.\ I.\ Ohba and  Prof.\ H.\ Nakazato for their helpful
comments and their continuous encouragement. We thank many discussions 
between Dr.\ H. Nakamura, Dr.\ T.\ Hirano and other members of 
Waseda Univ. high energy 
physics group. We also thank Dr.\ T.\ Peitzmann 
and Dr.\ S.\ Nishimura for 
 discussions about the recent WA98 results.  
Numerical analyses have been done with workstations of 
Waseda Univ. high energy physics group.  
%This work was partially supported by Kaken-shourei *****

\begin{Large}
{\bf Figure Caption}
\end{Large}

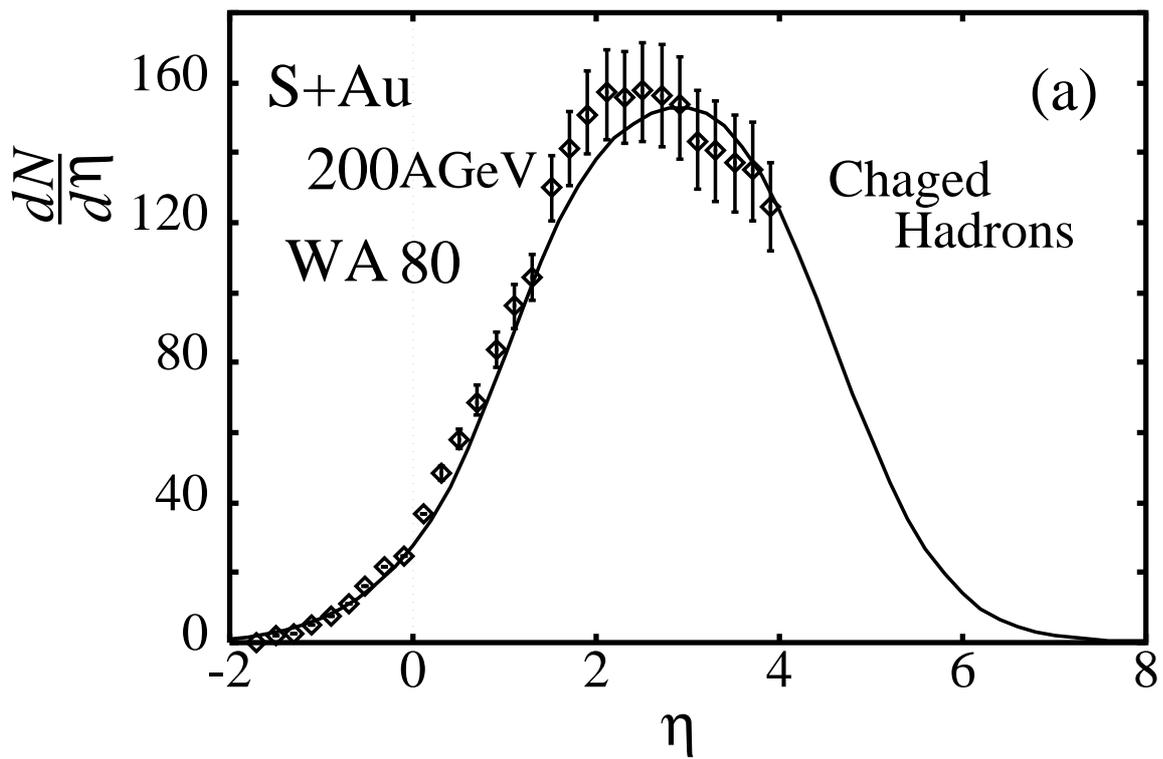
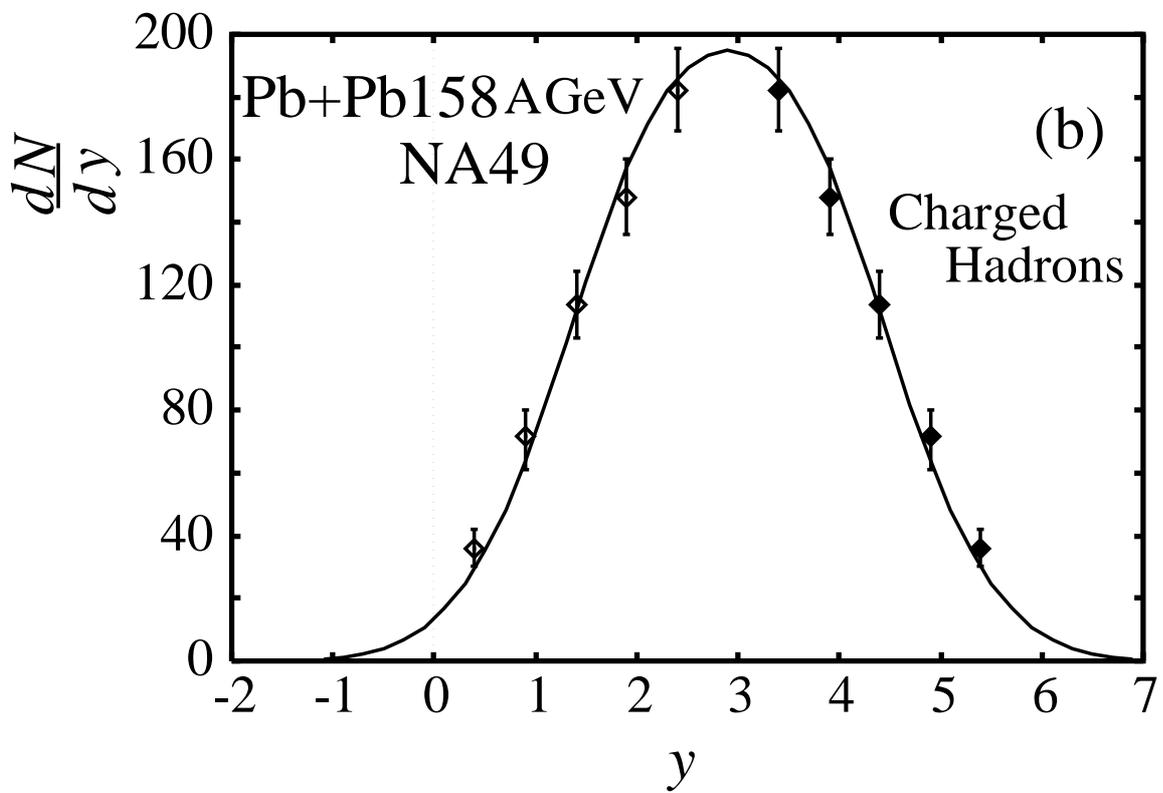
{\bf Figure1}:(a) Pseudo-rapidity distribution of charged hadrons in 
  S+Au 200 AGeV
 collision in comparison with the 
data of CERN WA80, and (b) rapidity 
 distribution of charged hadrons in Pb+Pb 158 AGeV 
collision in comparison with the data of CERN NA49.

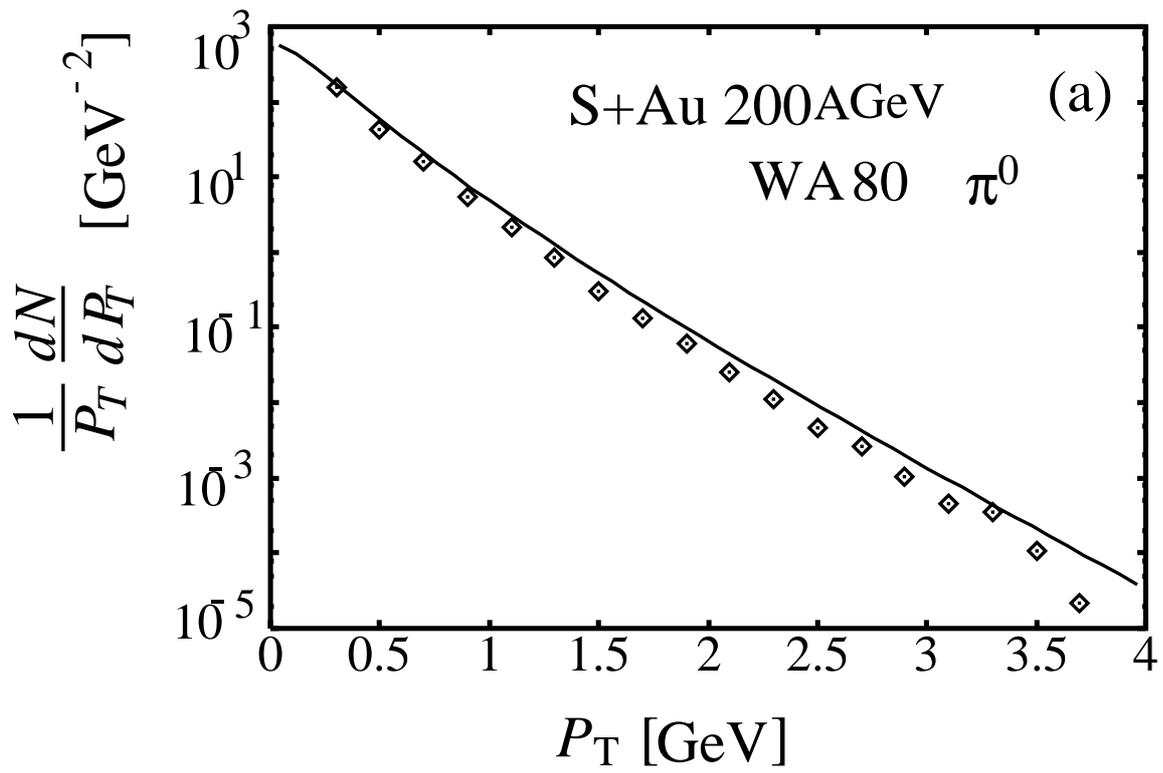
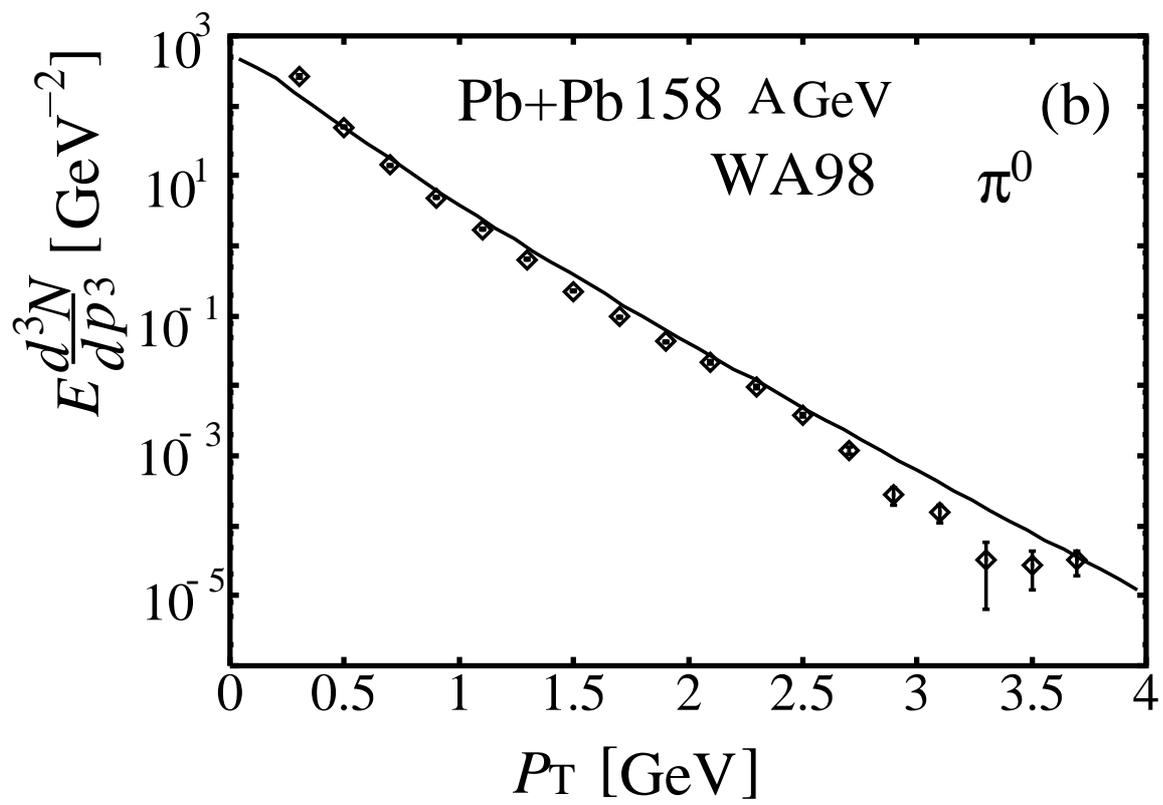
{\bf Figure2}:Transverse momentum distribution of 
neutral pions in S + Au 
collision 200 AGeV collision in comparison  with 
the data of CERN WA80(a),  Pb+Pb 158 AGeV collision 
in comparison with 
the data of CERN WA98(b).

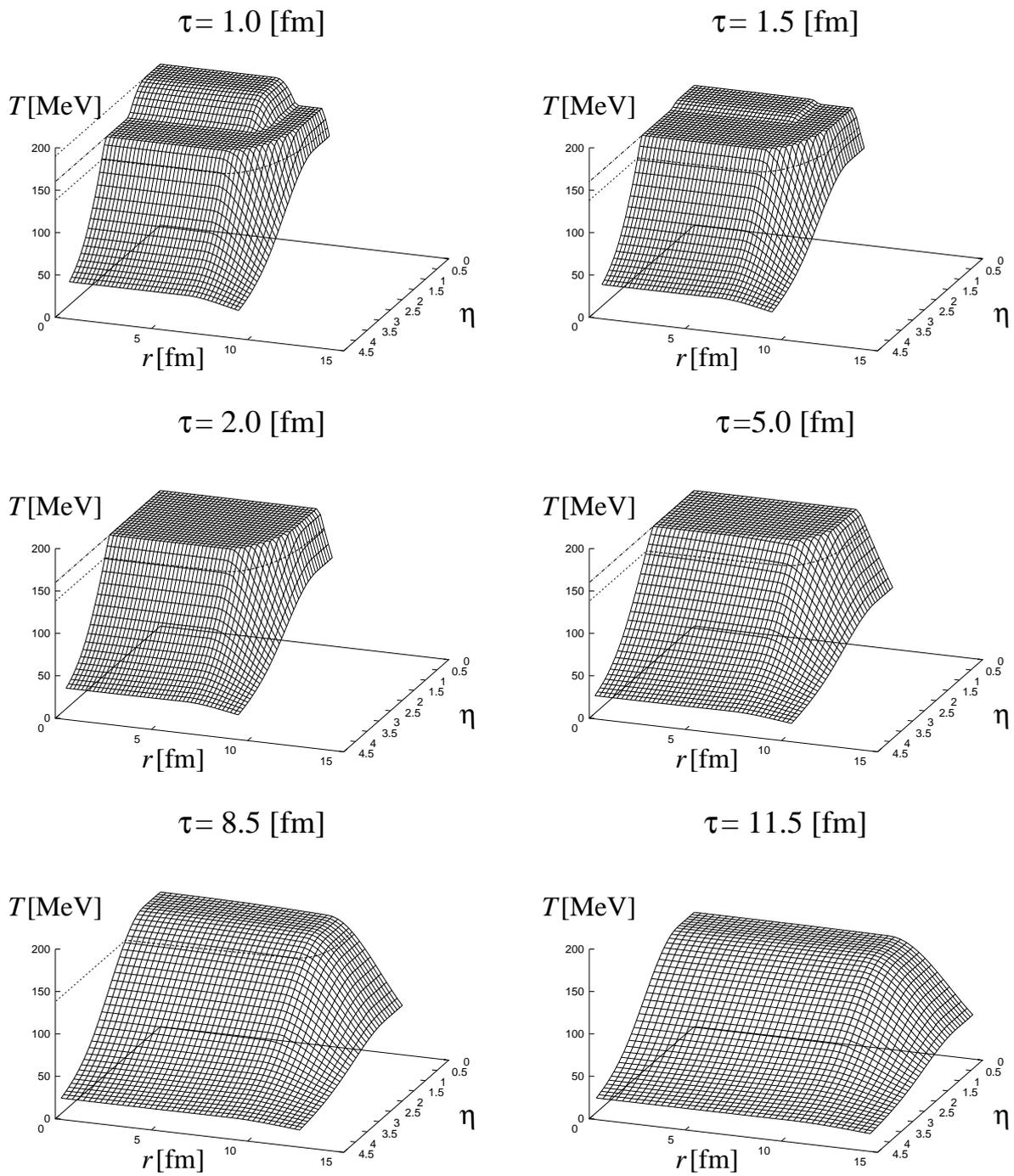
{\bf Figure3}:The space-time evolution of the temperature-distribution 
of the Pb+Pb 158 AGeV collision.  Initial temperature is 190 MeV, 
critical temperature is 160 MeV and freeze-out 
 temperature is 138 MeV.
The flat area corresponds to the mixed phase. 
From these graphs we can see that the QGP phase 
vanishes at $\tau$ = 2.0 fm
and the mixed phase vanishes at $\tau$ = 8.5 fm and 
at $\tau$ = 11.5 fm
the temperature becomes  lower than the freeze-out 
temperature everywhere in the fluid.

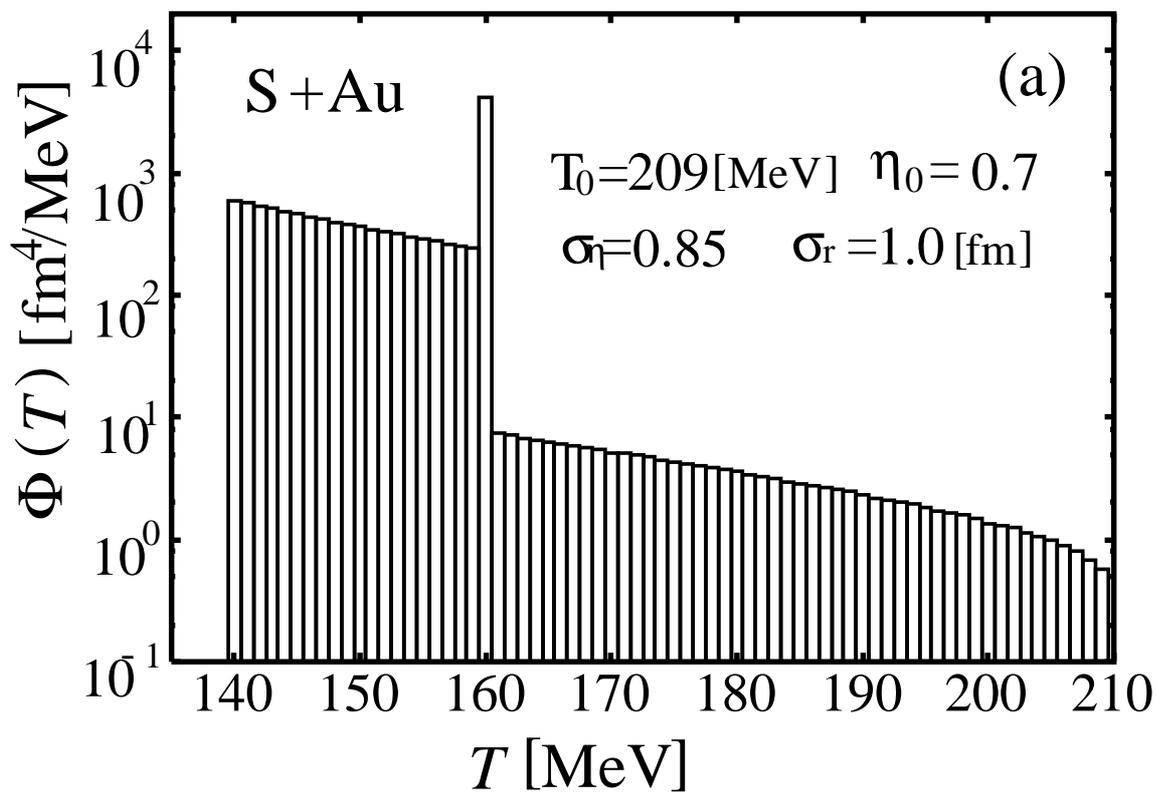
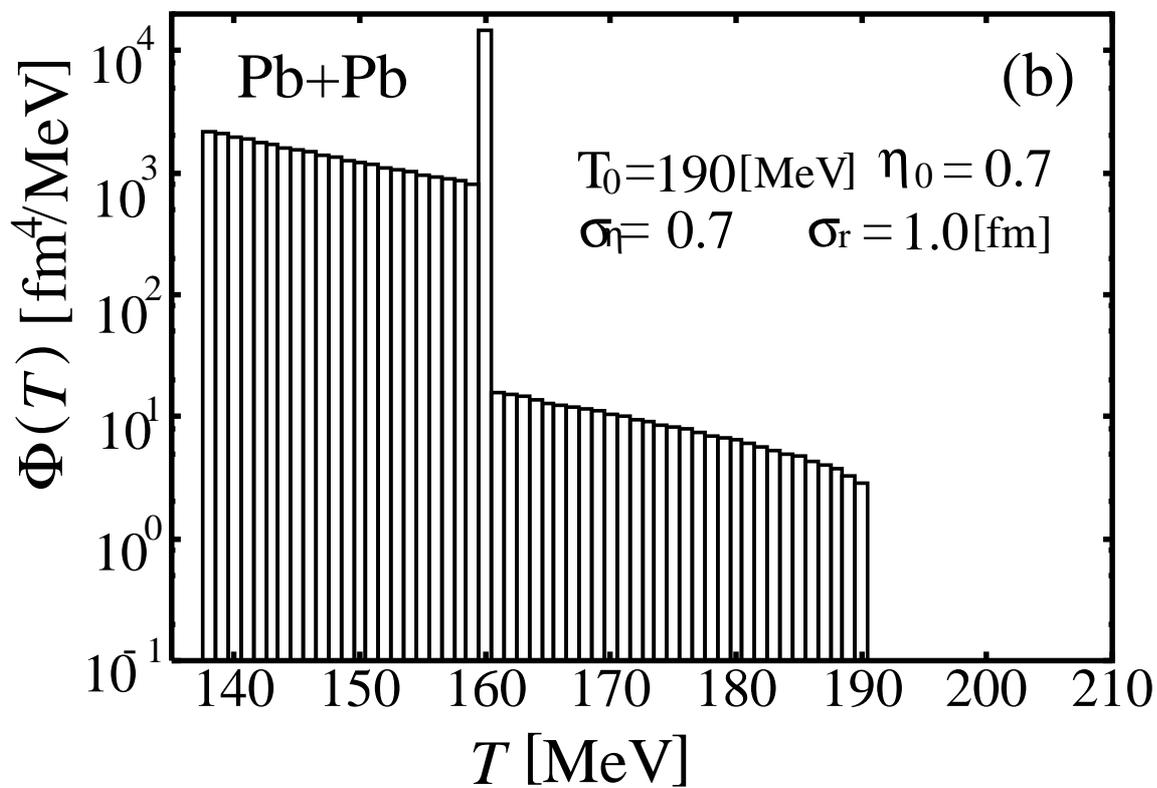
{\bf Figure4}:The temperature profile function of S+Au 200 
AGeV collision(a) and Pb+Pb 158 AGeV collision(b).
The sharp peak at critical temperature corresponds to
the huge space-time size of the mixed phase region
which is caused by the released latent heat.

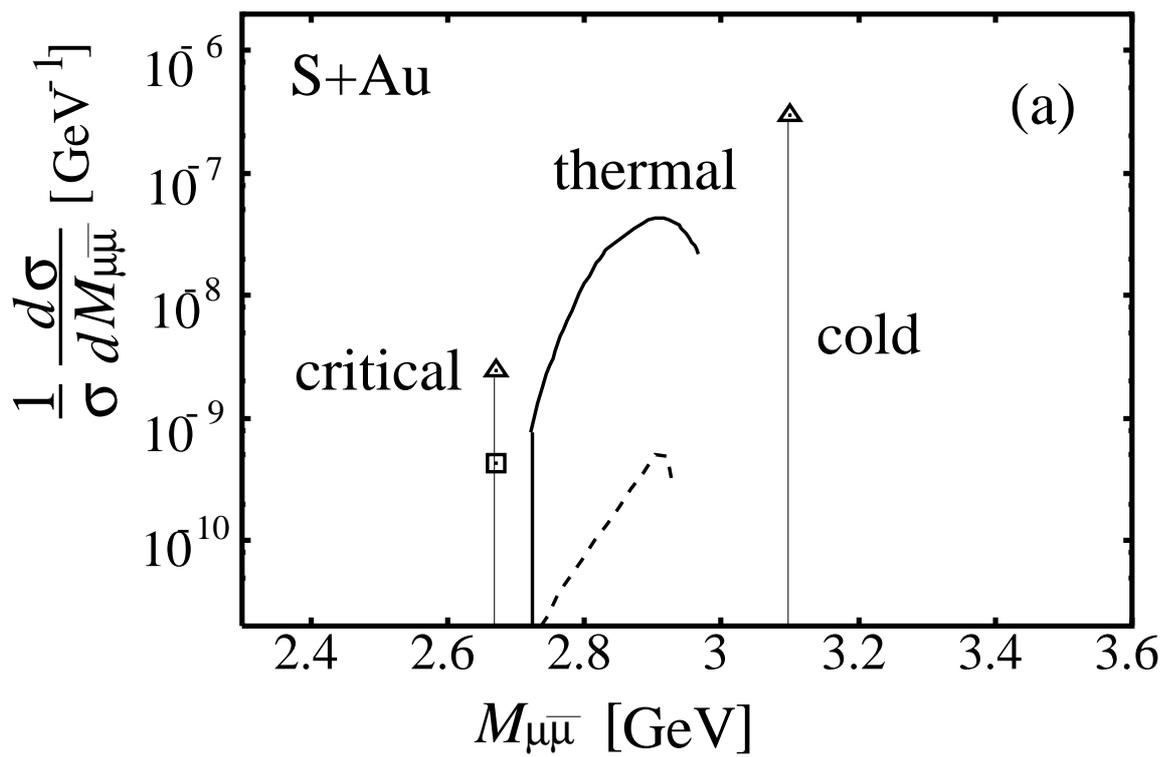
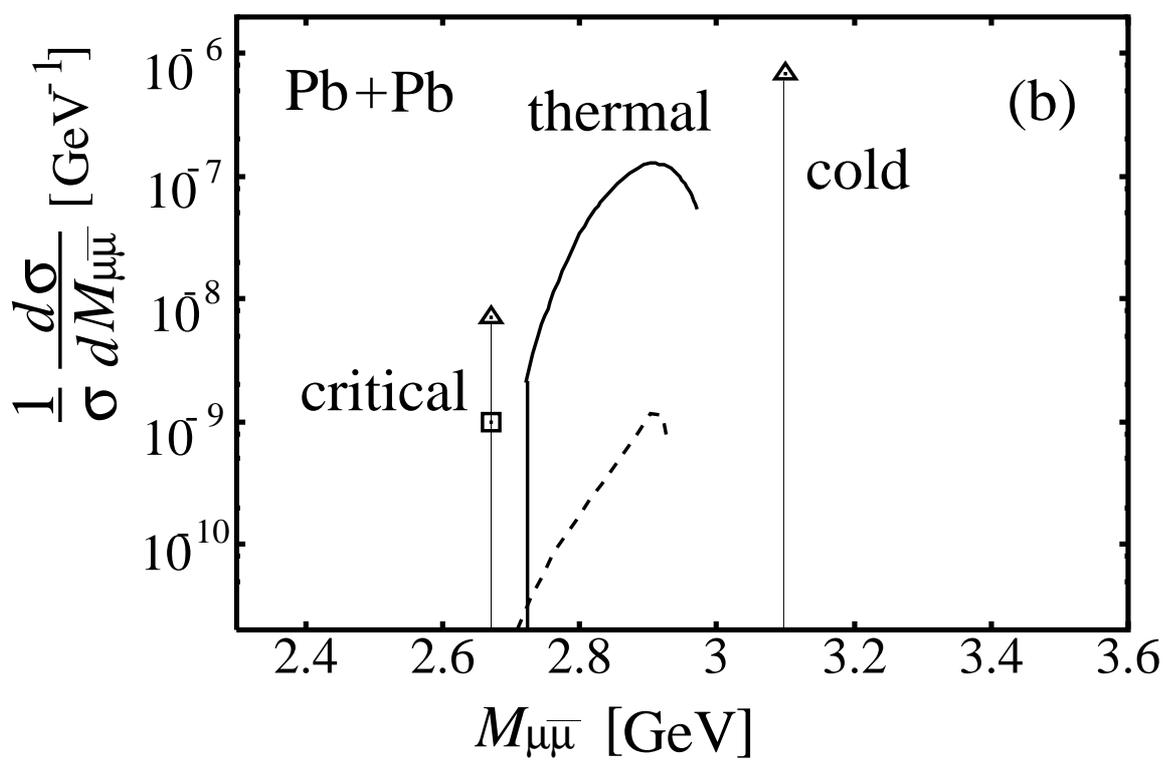
{\bf Figure5}:The invariant mass distribution of lepton pairs from 
mass-shifted $J/\psi$ in the hadronic phase of 
``critical",``thermal" and ``cold" in S+Au collision(a) and 
 Pb+Pb collision(b). 
The dashed line and a square point with a dot stand for 
the results obtained by Hioki et al.
In these figures, to make comparison clear, the results 
of Hioki et al. were so normalized as to give 
the same ``cold" contribution to our results.

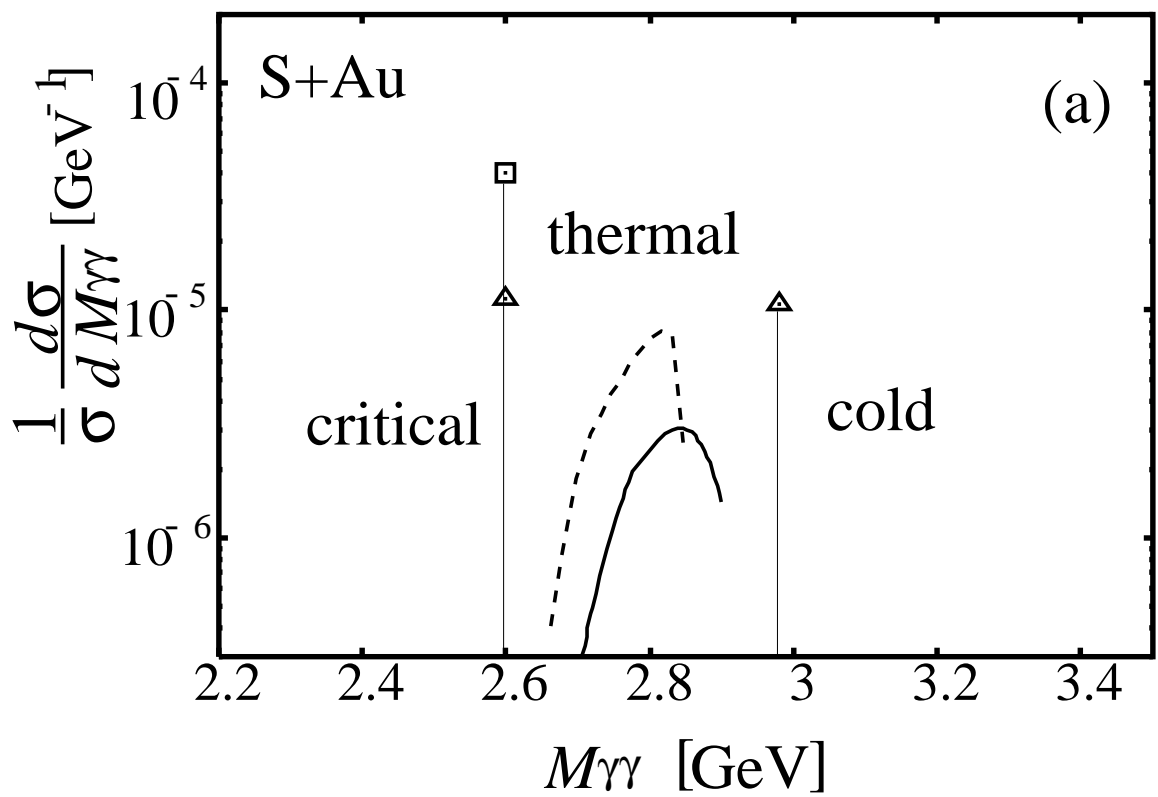
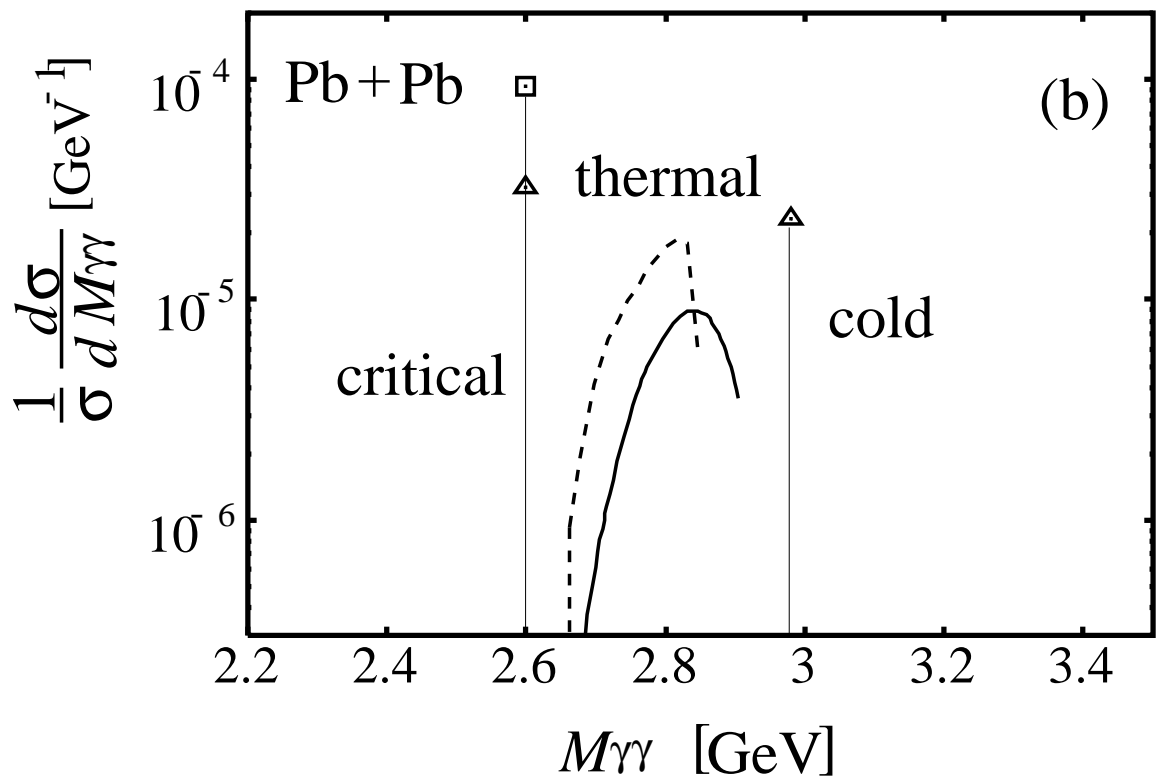
{\bf Figure6}:The invariant mass distribution of photon pairs from 
mass-shifted $\eta_{c} $ in the hadronic phase of 
``critical",``thermal" and ``cold" in S+Au collision(a) and Pb+Pb 
collision(b). 
The dashed line and a square point with a dot stand for
the results of Hioki et al.  In order to make comparison clear,   
the results of Hioki et al. were so normalized as to give 
the same ``cold" contribution to our  results.

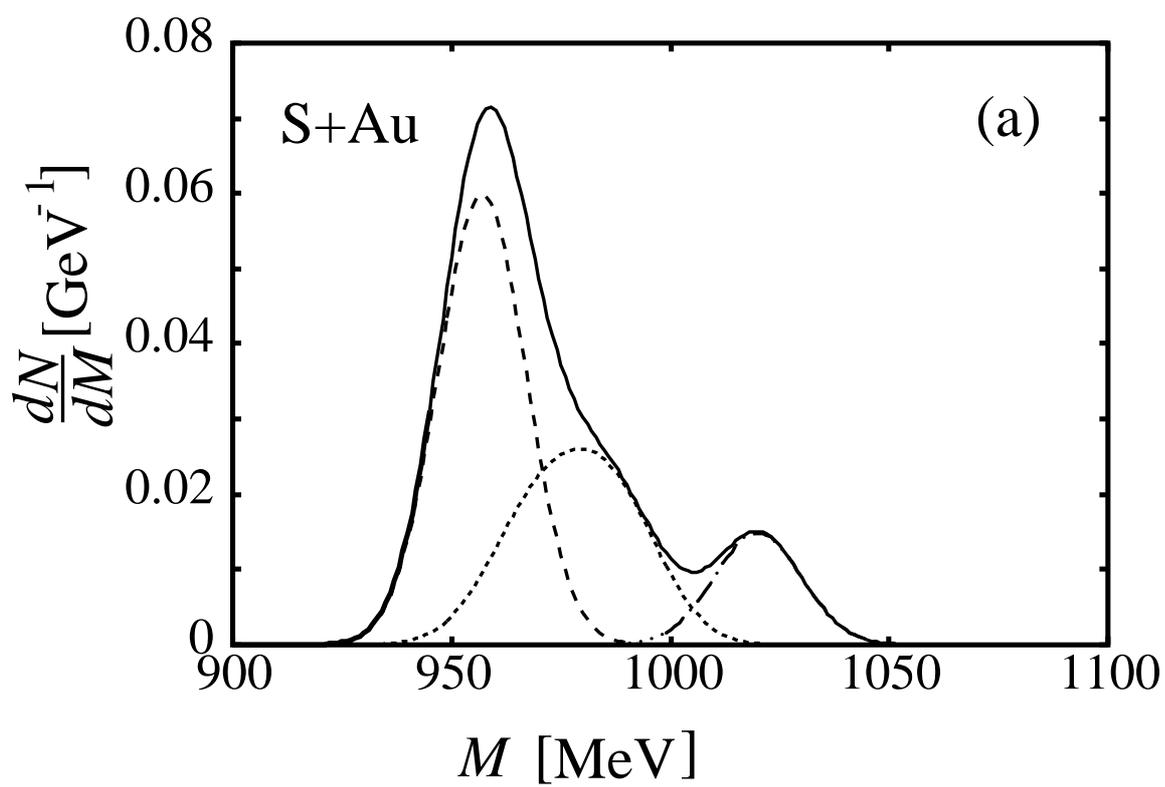
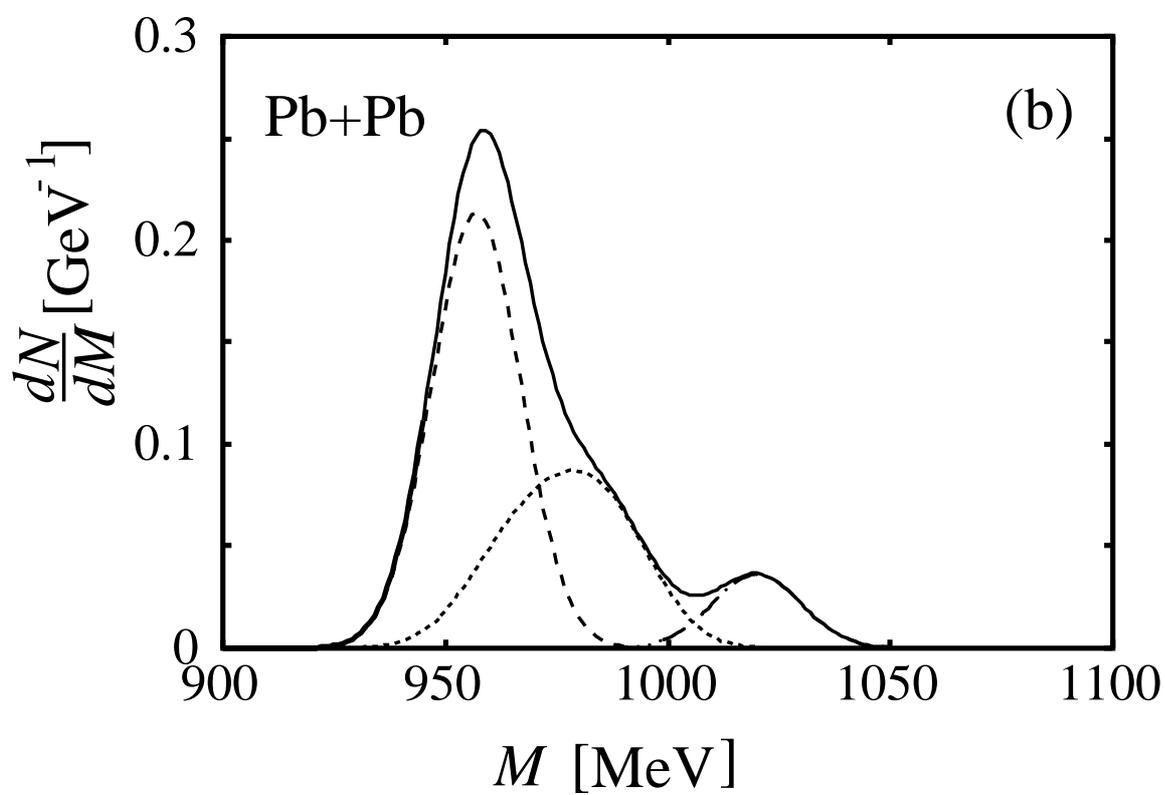
{\bf Figure7}:The invariant mass distribution of lepton pairs from 
mass-shifted $\phi$ in the central rapidity region
in S+Au collision(a) and Pb+Pb collision(b). 
The dashed line stands for the contribution from the mixed phase, 
the dotted line for the hadronic phase, the chained line for 
 the contribution  from the freeze-out hypersurface and
the solid line for the total contribution, respectively. 
 The peak at 1020 MeV corresponds to  the contribution 
from the freeze-out hypersurface.

\begin{thebibliography}{99}
\bibitem{QM}
See, for example, {\it Proceedings of Quark Matter '96},
{\em Nucl.~Phys.\/}{\bf A610} (1996)1c.

\bibitem{Kanaya}
K.\ Kanaya, {\em Nucl.\ Phys.\ }{\bf B}(Proc.\ Supple.){\bf 47}(1996)144; 
K.\ Kanaya in the Proceedings of the Third International Conference
of the Physics and AstroPhysics of Quark-Gluon Plasma( held at Jaipool, India (1997)).

\bibitem{Hydro}
J. D. Bjorken, Phys. Rev. {\bf D27}(1983)140.  

\bibitem{Date1}
Y.~Akase, S.~Dat{\'e}, M.~Mizutani, S.~Muroya, M.~Namiki and 
M.~Yasuda, {\em Prog.\ Theor.\ Phys.\ } {\bf 82}(1989)591.

\bibitem{Akase}
Y.~Akase, M.~Mizutani, S.~Muroya and M.~Yasuda,
{\em Prog.\ Theor.\ Phys.\ } {\bf 85} (1991)305;
S.~Muroya, H.~Nakamura and M.~Namiki,
{\em Prog.\ Theor.\ Phys.\ Suppl.\ }{\bf No.120}(1995)209.

\bibitem{Hirano}
T.\ Hirano, S.\ Muroya and M.\ Namiki {\em Prog.\ Theor.\ Phys.\ } {\bf 98}(1997)129

\bibitem{Alam}
J.\ Alam, S.\ Raha and B.\ Sinha, {\em Phys.\ Rep.\ }{\bf 273}(1996)243; 
J.\ Alam, D.\ K.\ Srivastava, B.\ Sinha and D.\ N.\ Basu,
{\em Phys.\ Rev.\ }{\bf D48}(1993)1117;
K.\ Kajantie, M.\ Kataja, L.\ McLerran and P.\ V.\ Ruuskanen, Phys.\ Rev. {\bf D34}(1986)811.

\bibitem{Sollfrank}
J.\ Sollfrank, P.\ Houvinen, M.\ Kataja, P.\ V.\ Ruuskannen, M.\ Prakash 
and R.\ Venugopalan, {\em Phys.\ Rev.\ }{\bf C55}(1997)391.

\bibitem{Ishii}
T.~Ishii and S.~Muroya, {\em Phys.\ Rev.\ } {\bf D46}(1992)5156. 

\bibitem{Arbex}
N.~Arbex; 
U.~Ornik, M.~Pl{\"u}mer, A.~Timmermann and
R.~M.~Weiner, {\em Phys. Lett.}{\bf B345}(1995)307.
U.\ Ornic, M.\ Pl{\"u}mer, B.R.\ Schlei, D.\ Strottman, and R.M.\ Weiner,
{\em Phys.\ Rev.\ }{\bf D54}(1996)1381.

\bibitem{Hioki}
S.\ Hioki, T.\ Kanki and O.\ Miyamura, Phys.\ Lett.\ {\bf B 261}(1991)5; 
T.\ Hashimoto, K.\ Hirose, T.\ Kanki and O.\ Miyamura, Z.\ Phys.\ {\bf C38}
(1988)251. 

\bibitem{Asakawa}
M.\ Asakawa and C.\ M.\ Ko, Phys.\ Lett.\ {\bf B 322}(1994)33.

\bibitem{92WA80}
R.~Albrecht et al. : WA80, {\em Z.~Phys.\/ }{\bf C55}(1992)539.

\bibitem{94WA80}
R.~Santo et al. : WA80, {\em Nucl.\ Phys.\ }{\bf A566}(1994)61c.

\bibitem{96NA49}
S.~V.~Afanasiev et al. :NA49, {\em Nucl.\ Phys.\ }{\bf A610}(1996)188c.

\bibitem{WA98}
M.~Aggarwal et al. : WA98, {\em Nucl.\ Phys.\ }{\bf A610}(1996)200c;
T.~Peitzmann, private communication.

\end{thebibliography}
\end{document}